\documentclass[aps,preprint,amsmath,amssymb,superscriptaddress,footinbib]{revtex4-1}
\usepackage{graphicx,color,multirow,bm,braket} %lineno
\usepackage[colorlinks,allcolors=black]{hyperref}

\def\bj{{\bf j}}
\def\br{{\bf r}}
\def\bv{{\bf v}}
\def\kb{k_\textup{B}}

\renewcommand{\thetable}{\arabic{table}}
\makeatletter
\renewcommand\@make@capt@title[2]{%
\@ifx@empty\float@link{\@firstofone}{\expandafter\href\expandafter{\float@link}}%
{\textbf{#1}}\@caption@fignum@sep#2\quad}%
\renewcommand{\fnum@figure}{\textbf{Fig.~\thefigure}}
\renewcommand{\fnum@table}{\textbf{Table~\thetable}}
\makeatother

\begin{document}

%%% TITLE <= 15 words
\title{Thermal transport and phase transitions of zirconia by on-the-fly 
machine-learned interatomic potentials}

\author{Carla Verdi}
\email{carla.verdi@univie.ac.at}
\affiliation{University of Vienna, Faculty of Physics, Computational Materials Physics, 1090 Vienna, Kolingasse 14-16, Austria} %
\author{Ferenc Karsai}
\affiliation{VASP Software GmbH, 1090 Vienna, Sensengasse 8, Austria} %
\author{Peitao Liu}
\affiliation{VASP Software GmbH, 1090 Vienna, Sensengasse 8, Austria} %
\author{Ryosuke Jinnouchi}
\affiliation{Toyota Central Research and Developments Laboratories, Inc., Aichi 480-1192, Japan}
\author{Georg Kresse}
\affiliation{University of Vienna, Faculty of Physics, Computational Materials Physics, 1090 Vienna, Kolingasse 14-16, Austria} %
\affiliation{VASP Software GmbH, 1090 Vienna, Sensengasse 8, Austria} %

%%% ABSTRACT <=150 words
\begin{abstract}
Machine-learned interatomic potentials enable realistic finite temperature
calculations of complex materials properties with first-principles accuracy. 
It is not yet clear, however, how accurately they describe anharmonic 
properties, which are crucial for predicting the lattice thermal 
conductivity and phase transitions in solids and, thus, shape their 
technological applications.
Here we employ a recently developed on-the-fly learning technique based on
molecular dynamics and Bayesian inference in order to generate an interatomic
potential capable to describe the thermodynamic properties of zirconia, an
important transition metal oxide. This machine-learned potential accurately
captures the temperature-induced phase transitions below the melting point.
We further showcase the predictive power of the potential by calculating the
heat transport on the basis of Green-Kubo theory, which allows to account for
anharmonic effects to all orders. 
This study indicates that machine-learned potentials trained on 
the fly offer a routine solution for accurate and efficient simulations of 
the thermodynamic properties of a vast class of anharmonic materials.
\end{abstract}

\maketitle

%%% MAIN TEXT (excluding abstract, methods, refs, figure legends) <5000 words;
%%% up to 10 display items (figures, tables)
\section{Introduction} \label{sec:intro}

The atomistic modelling of thermodynamic properties of materials at finite
temperature and realistic conditions is of fundamental importance for a multitude
of technological applications, from photovoltaics to optoelectronic devices, as
well as for advancing our understanding of the physical properties of matter.
This represents a huge challenge for modern atomistic simulations. Molecular
dynamics (MD) and Monte Carlo techniques offer powerful tools that, when used
in conjunction with \textit{ab~initio} (AI) theories, allow us to achieve the desired
accuracy and predictive power~\cite{Marx-Hutter}. First-principles based methods,
however, suffer from serious limitations. For instance, the description of many
thermodynamic properties, even in simple crystals, requires accessing time
and size scales that are far beyond the capabilities of state-of-the-art methods.

Two of the most challenging, yet central, properties are temperature-induced
structural phase transitions in solids and the transport of heat, which in
semiconductors and insulators mainly stems from the lattice vibrations. Both
can be extracted directly from MD simulations~\cite{Tuckerman}. In the case
of heat transport, this can be obtained from equilibrium MD calculations on
the basis of the Green-Kubo (GK) theory~\cite{Baroni2018}. This method, based on
the fluctuation-dissipation theorem, is exact at sufficiently high temperatures
where nuclear quantum effects are negligible and is thus superior to lattice
dynamics approaches based on the Boltzmann transport equation (BTE) in crystalline
systems \cite{Lindsay2018}. AI calculations of thermal transport based on GK
theory, however, have only recently become possible, after a fundamental
issue related to the ill-definition of the microscopic heat flux was
resolved~\cite{Baroni2016}. Recently, two new theories of thermal transport
have also been developed. These are capable of treating crystalline and
amorphous systems on the same footing and reduce to the BTE in the limit of
periodic solids~\cite{Mauri2019,Baroni2019}. Yet, the GK approach carries the
advantage that it accounts exactly for anharmonicity to all orders, while
intrinsically lending itself to a unified description of ordered and disordered
solids, as well as liquids. Nevertheless, the long simulation times and large
supercell sizes required in order to achieve converged results are still
prohibitive for AI calculations. While some approaches have been devised to
overcome this problem~\cite{Carbogno2017,Baroni2017}, they are not sufficient
to straightforwardly enable calculations of heat transport from first-principles
in any type of solid.

MD simulations based on (semi)empirical interatomic potentials are not affected
by these limitations, as the computing time for each MD step is generally more
than five orders of magnitude lower than with first-principles methods.
Interatomic potentials, however, are lengthy and cumbersome to develop, have
very limited flexibility and transferability, and generally do not offer sufficient
accuracy~\cite{Bokdam2018} even in the case of simple systems~\cite{Galli2012}.
Machine-learned force fields (MLFFs) have recently emerged as an ideal solution
to drastically augment the length and time scales accessible to MD simulations
while still retaining first-principles accuracy~\cite{Behler2016,Bartok2017}.
Machine-learning models based on mapping the local environment around each atom
in the system onto a set of descriptors allow to train force fields that reproduce
first-principles energies, forces, and stresses with very small errors.
A key issue is the construction of the training sets. To avoid expensive
and somewhat manual procedures for building a training database, so-called active
learning schemes are becoming increasingly popular~\cite{DeVita2015,Shapeev2017,
Jinnouchi2019b,Sivaraman2020}. In particular, a number of algorithms have been put
forward in order to generate potentials on the fly, relying on query strategies to
allow the machine to judge whether new structures need to be added to the training
dataset. An active learning strategy where structures are generated on the fly
during MD simulations, combined with Bayesian inference to estimate the
uncertainty of the machine-learning model, has been shown to be especially
flexible and efficient, and it can be seamlessly integrated into existing
first-principles codes~\cite{Jinnouchi2019b,Kozinsky2020}.

The success of machine-learned potentials used in combination with MD simulations
has been demonstrated in a variety of applications, from the study of phase
transitions in halide perovskites~\cite{Jinnouchi2019a} and lithium manganese
oxide spinels~\cite{Behler2020} to thermal conductivities in solid and amorphous
silicon~\cite{Qian2019} and the CoSb$_3$ skutterudite~\cite{Shapeev2019}. 
For any MLFF, however, it is extremely 
challenging to simultaneously describe vibrational properties, dynamical changes 
of lattice parameters during phase transitions, and anharmonic effects relevant 
for transition temperatures as well as thermal conductivity. In fact, none of 
Refs.~\cite{Jinnouchi2019a,Behler2020,Qian2019,Shapeev2019} show the ability 
of MLFFs to describe all these different thermodynamic properties simultaneously. 
Harmonic vibrational properties are already a stringent test of the quality of ML 
potentials, especially when considering materials with soft phonon modes and 
different polymorphs~\cite{George2020}. However, the description of phonon 
dispersion relations requires the acquisition of only a two-body term, the 
harmonic force constants. Anharmonicity, on the other hand, requires the 
development of a surrogate model for tiny, subtle energy differences on the 
highly corrugated harmonic energy surface. The MLFF must capture at least 
third-order anharmonic force constants, the coupling between phonons and 
lattice distortions, and, especially at high temperature, even higher-order 
anharmonic force constants. %four-particle interactions. 
For example, calculating just the third-order terms easily requires more than 
a thousand first-principles calculations.
Whether machine-learned potentials can describe all these effects with 
satisfactory accuracy has  been insufficiently investigated.

In this work we address this gap by focusing on one paradigmatic 
example of anharmonic systems, zirconia (ZrO$_2$), and we present a procedure to
accurately study its thermodynamic properties using MLFFs trained 
on the fly. This allows for a semi-automated creation of a reference database. 
ZrO$_2$ is a transition metal oxide of tremendous importance in a wide variety of
applications, including thermal barrier coatings, solid oxide fuel cells, and
biomedical implants~\cite{Clarke2003,Denry2008}. Pure ZrO$_2$ exhibits low thermal
conductivity due to strong anharmonic phonon scattering~\cite{Mevrel2004}, thus
representing an ideal testbed for the investigation of thermal transport via GK
theory and a challenging system for building accurate MLFFs.

\begin{figure*}[t] \begin{center}
 \includegraphics[width=\textwidth]{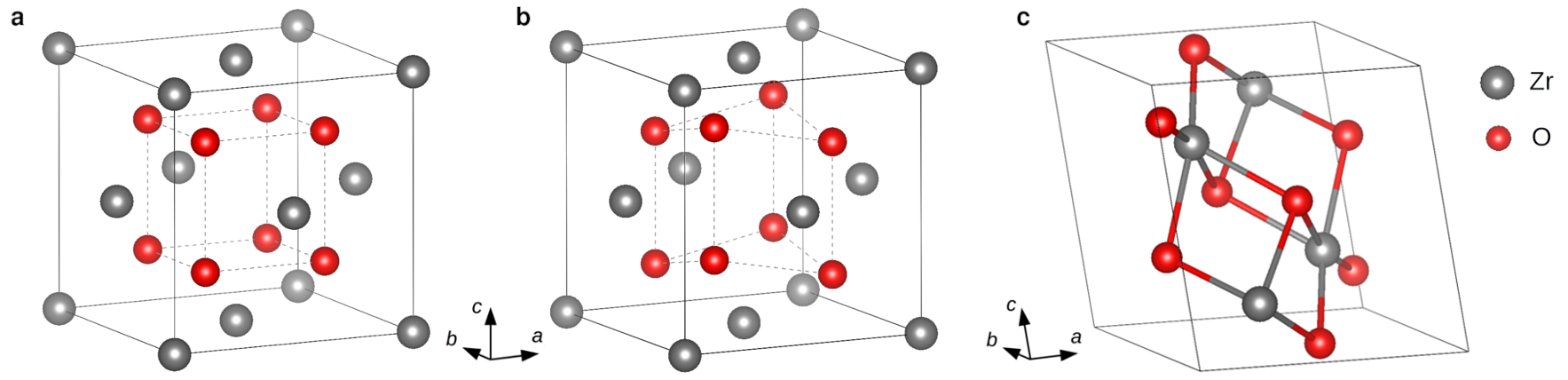}
 \caption{ \label{fig1}
 \textbf{Crystal structures of zirconia.}
  Ball-and-stick model of the three structural phases of ZrO$_2$: \textbf{a}
  cubic, \textbf{b} tetragonal, and \textbf{c} monoclinic. The 12-atom
  conventional cells are represented, and the Zr-O bonds are shown only
  in the monoclinic structure. }
 \end{center} \end{figure*}

At ambient pressure, ZrO$_2$ is known to have three structural
phases~\cite{Subbarao1974,Aldebert1985} shown in Fig.~\ref{fig1}. At
high temperature the cubic fluorite structure (space group $Fm\bar3m$)
is thermodynamically stable and it transforms to a tetragonal phase (space
group $P4_2/nmc$) at about 2570~K, with the oxygen sublattice shifting up
or down along the $c$ direction (Fig.~\ref{fig1}b) accompanied by a slight
tetragonal distortion.
Around 1400~K the structure undergoes a tetragonal to monoclinic (space group
$P2_1/c$) phase transition. This transition is of martensitic type, involving
a complex cooperative motion of both the Zr and O sublattices accompanied by a
lattice shear in the $ac$ plane and a volume increase. The phase diagram of
zirconia is extremely important for its technological applications, as for
instance the volume change associated to the tetragonal to monoclinic transition
can lead to mechanical degradation. Moreover, the tetragonal and cubic structures
can be stabilized at lower temperatures by introducing dopants such as yttria,
ceria, and other oxides. Stabilized zirconia is one of the most important ceramic
materials due to its combination of high strength and toughness at room
temperature~\cite{Pascoe1975,Chevalier2020}.
The very high temperature of the cubic to tetragonal transition makes it
challenging for experiments to understand its nature and dynamics~\cite{Kisi1998},
while computational studies report conflicting results~\cite{Finnis2001,
Schelling2001,Carbogno2014}. On the opposite, the first-order tetragonal to
monoclinic transition has been extensively characterized, but theoretical
calculations are limited to approximate treatments such as the quasi-harmonic
approximation~\cite{Parlinski2005,Tanaka2005}.

Here we use a recently developed on-the-fly training scheme~\cite{Jinnouchi2019b}
for building a machine-learned interatomic potential that describes all three
polymorphs of zirconia with near first-principles accuracy. In contrast, no
empirical interatomic potential is capable of seamlessly describing all phases,
nor to reach such accuracy~\cite{Schelling2001}. We validate %benchmark
the MLFF against the results from first-principles density-functional theory
(DFT) calculations and use it to determine the phase transitions and heat
transport in ZrO$_2$ at ambient pressure, accounting for anharmonic 
effects to all orders. The agreement with experiments
demonstrates that machine-learned interatomic potentials constructed on the fly
offer an excellent, highly automated solution for accurate and realistic
predictions of anharmonic properties in complex materials.

%%% RESULTS (DISCUSSION generally a separate section) should have subheadings
\section{Results} \label{sec:results}

\subsection{Machine-learned potential}

In order to machine learn an interatomic potential for zirconia we used the
kernel-based machine-learning model introduced in 
Refs.~\cite{Jinnouchi2019b,Jinnouchi2020}. 
Details of the model are presented in the Methods section, including the
on-the-fly force field generation scheme that we adopted. This scheme is
fully integrated in the VASP code~\cite{Kresse1993,Kresse1996,Kresse1996b}.

As we aim to compute the thermodynamic properties of a strongly anharmonic material 
up to very high temperatures, the configurational phase space should be sampled as 
broadly and as efficiently as possible. Our on-the-fly training strategy, described 
in the Methods, is well suited for this, allowing us to seamlessly sample 
an ample configuration space including large thermal fluctuations up to 2800~K 
during MD simulations. In fact, training on the fly enables us to 
skip most \textit{ab~initio} steps and perform \textit{ab~initio} calculations 
only when required, that is, when the estimated Bayesian error of the MLFF is large. 
As a result, only 592 structures consisting of 96-atom supercells were automatically 
selected for performing \textit{ab~initio} calculations from a trajectory over 300~ps 
long, and were included in the final training dataset. Nevertheless, the resulting 
MLFFs are very accurate.

 \begin{table}[b] \begin{center}
 \caption{ \label{tab:errors}
 RMSE in the energies, forces, and stress tensors predicted by the MLFF for the
 training and test dataset (the latter in parenthesis). The RMSE in the predicted
 phonon energies are also reported (in meV). All results are shown for the MLFFs
 trained both using BLR and SVD.}
 \begin{ruledtabular}
 \begin{tabular}{lccc}
 \multicolumn{4}{c}{Training and validation errors} \\
 & Energy (meV per atom) & Force (eV \AA$^{-1}$) & Stress (kbar) \\
 \colrule
 BLR & 1.70 (1.78) & 0.13 (0.13) & 1.87 (1.89) \\
 SVD & 1.45 (1.42) & 0.11 (0.11) & 1.67 (1.76) \\
 \colrule
 \multicolumn{4}{c}{Phonon errors} \\
 & Cubic & Tetragonal & Monoclinic \\
 \colrule
 BLR & 2.73 & 0.96 & 0.68 \\
 SVD & 1.59 & 0.80 & 0.55 \\
 \end{tabular}
 \end{ruledtabular}
 \end{center} \end{table}
 
The root-mean-square errors (RMSE) in the energies, forces, and stress
tensors predicted by our MLFFs for the training dataset are reported in
%Table~\ref{tab:errors}. 
Table~1. In particular, we used two distinct approaches to
obtain the regression coefficients typical of kernel-based methods, namely
Bayesian linear regression (BLR) and singular value decomposition (SVD, see
Methods). While BLR is formally similar to the standard ridge 
regression methods, SVD does not introduce any regularization parameters. 
Here we compare the accuracy of two different MLFFs obtained using BLR and 
SVD and find that SVD yields consistently lower errors than BLR. 
Overall, we observe that our MLFFs achieve 
lower errors than the neural-network potential developed in Ref.~\cite{Kitchin2018}, 
which has an RMSE in the energy of 7.9~meV per atom, and whose training required
a more manual setup of many more DFT calculations. Although the training dataset of
Ref.~\cite{Kitchin2018} included additional structures containing oxygen
vacancies, here we considered temperatures up to 2800~K, which induce strong
fluctuations that are considerably challenging to reproduce.

 \begin{figure} \begin{center}
 \includegraphics[width=0.5\columnwidth]{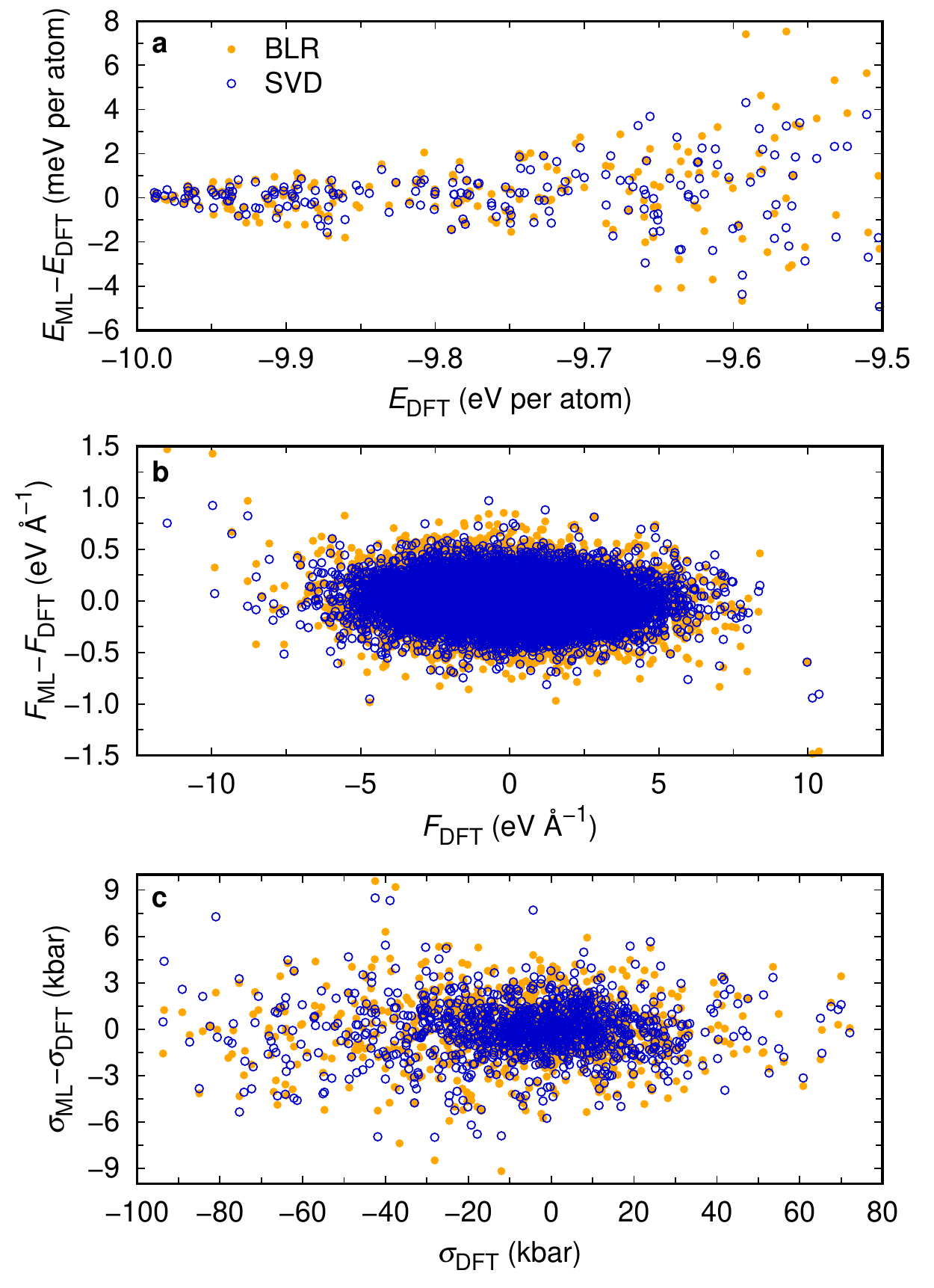}
 \caption{ \label{fig:errors}
 \textbf{Errors of the machine-learned force fields.}
 Difference between the \textbf{a} energies, \textbf{b} atomic forces, and
 \textbf{c} stress components predicted by the MLFFs (in orange using BLR,
 in blue using SVD for the regression) and the DFT ones for the structures in
 the test dataset.}
 \end{center} \end{figure}

To validate the MLFFs, energies, forces and stress tensors were calculated
for a test set consisting of 200 structures and compared to the DFT ones,
as shown in Fig.~\ref{fig:errors}. The structures were selected randomly
from an MD simulation where monoclinic ZrO$_2$ was heated between 300~K and
2800~K. The largest errors correspond to the high-energy structures, that is
the cubic phase at elevated temperatures. In this regime, the MLFF is less
accurate due to the increased thermal vibrations. The RMSE for this test set,
reported in %Table~\ref{tab:errors}
Table~1, are very close to the ones obtained for
the training set. This shows that the MLFFs were not overfitted, even in the
SVD case where no regularization was used.

\subsection{Structural and vibrational properties}

 \begin{table} \begin{center}
 \caption{ \label{tab:latt-par}
 Structural parameters for all three phases of ZrO$_2$ and energy differences
 between phases (per 12-atom cell). In the tetragonal phase, $d_z$ is the
 displacement of the oxygen atoms along the $c$ direction with respect to
 the cubic structure (in units of $c$), while $\beta$ is the angle between the
 lattice vectors $a$ and $c$ in the monoclinic phase. The results from the MLFF
 (both using BLR and SVD) are compared to the PBEsol calculations and the
 experimental data taken from Refs.~\cite{Aldebert1985,Howard1988,Catlow1994}. }
 \begin{ruledtabular}
 \begin{tabular}{lcccc}
 & PBEsol & BLR & SVD & Expt. \\
 \hline
 Cubic & & & & \\
 a (\AA) & 5.069 & 5.069 & 5.069 & 5.090 \\
 $E_\textup{c}-E_\textup{t}$ (eV) & 0.230 & 0.223 & 0.222 & \\
 \hline
 Tetragonal & & & & \\
 $a$ (\AA) & 5.074 & 5.073 & 5.075 & 5.050 \\
 $c$ (\AA) & 5.173 & 5.178 & 5.184 & 5.182 \\
 $d_z$ $(c)$ & 0.047 & 0.047 & 0.048 & 0.049 \\
 $E_\textup{t}-E_\textup{m}$ (eV) & 0.317 & 0.325 & 0.322 & \\
 \hline
 Monoclinic & & & & \\
 $a$ (\AA) & 5.133 & 5.130 & 5.136 & 5.151 \\
 $b$ (\AA) & 5.215 & 5.218 & 5.213 & 5.212 \\
 $c$ (\AA) & 5.300 & 5.296 & 5.306 & 5.317 \\
 $\beta$ (deg) & 99.56 & 99.67 & 99.63 & 99.23 \\
 \end{tabular}
 \end{ruledtabular}
 \end{center} \end{table}
 
The calculated zero-temperature lattice parameters for the conventional 12-atom
unit cells are summarized in %Table~\ref{tab:latt-par}
Table~2 and compared to the
experimental data \cite{Aldebert1985,Howard1988,Catlow1994} (the cubic and
tetragonal experimental lattice parameters are extrapolated to zero temperature).
The energy differences between the three phases are also shown in the table.
AI calculations based on the PBEsol exchange-correlation functional~\cite{Perdew2008} describe very accurately the relaxed structures
compared to experiment, almost on par with benchmark many-electron
calculations~\cite{Mittendorfer2020}. The MLFF results are in excellent agreement
with the AI ones, the differences being less than 0.2\% for the lattice parameters
and less than 1~meV per atom for the energies. Note that here BLR and SVD perform
equally well.

 \begin{figure*}[t] \begin{center}
 \includegraphics[width=\textwidth]{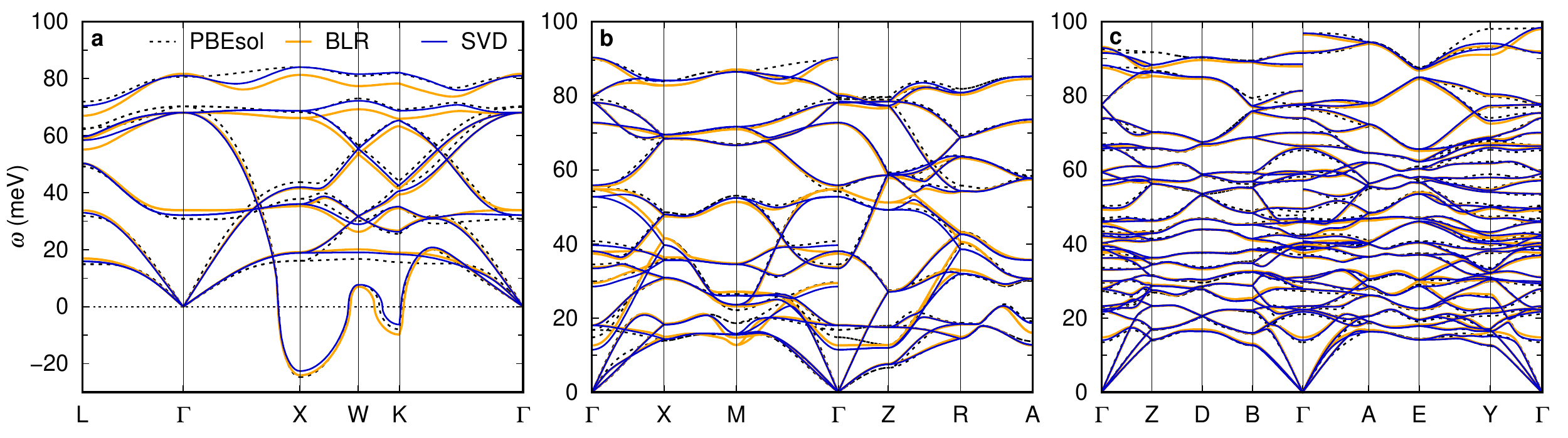}
 \caption{ \label{fig:phonons}
 \textbf{Prediction of phonon frequencies.}
 Phonon dispersions of \textbf{a} cubic, \textbf{b} tetragonal, and
 \textbf{c} monoclinic ZrO$_2$ along high-symmetry lines in the Brillouin
 zone. The results obtained using the MLFFs trained with BLR (orange lines)
 and SVD (blue lines) for the regression are compared with the AI ones
 calculated using PBEsol (dashed black lines). }
 \end{center} \end{figure*}

The MLFFs were then used to calculate the interatomic force constants and
dynamical matrices according to the finite displacement method in a supercell
containing 96 atoms. To account for the long-range dipole-dipole interactions,
the nonanalytic contribution to the dynamical matrix was treated using the
standard method of Ref.~\cite{Gonze1997}.
As illustrated in Fig.~\ref{fig:phonons}, the MLFFs reproduce accurately the AI
phonon dispersion relations for each phase, despite the MLFFs being trained
using a semi-automated on-the-fly procedure, without systematically distorting
the ideal  structures around the ground state to reproduce phonon properties.
Some discrepancies in the dispersion of the highest-energy optical phonon modes
can be attributed to finite-size effects. Note that the MLFF trained using BLR
is slightly less accurate than using SVD, as seen in particular for the highest
optical phonon branches in the cubic phase. The RMSE of the phonon energies
reported in %Table~\ref{tab:errors} 
Table~1 confirms this trend. It is not surprising
that the largest errors appear in the cubic phase: since it is 
unstable at low temperature, the training database contains hardly any 
configuration corresponding to this unstable, ideal structure. It is thus harder 
for the MLFF to reproduce the 0~K properties of the ideal undistorted cubic phase. 
Remarkably, both MLFFs accurately reproduce the imaginary 
soft phonon modes in cubic ZrO$_2$, which is 
essential in order to quantitatively capture phase transitions. 
In fact, such modes are generally associated with lattice 
instabilities, leading to structural phase transformations in the material. 
In ZrO$_2$, the soft phonon at the X point in the cubic phase, which involves 
a tetragonal distortion of the oxygen sublattice, is linked to the cubic to 
tetragonal phase transition~\cite{Parlinski1997}. 

\subsection{Temperature-induced phase transitions}

\begin{figure*}[t] \begin{center}
 \includegraphics[width=\textwidth]{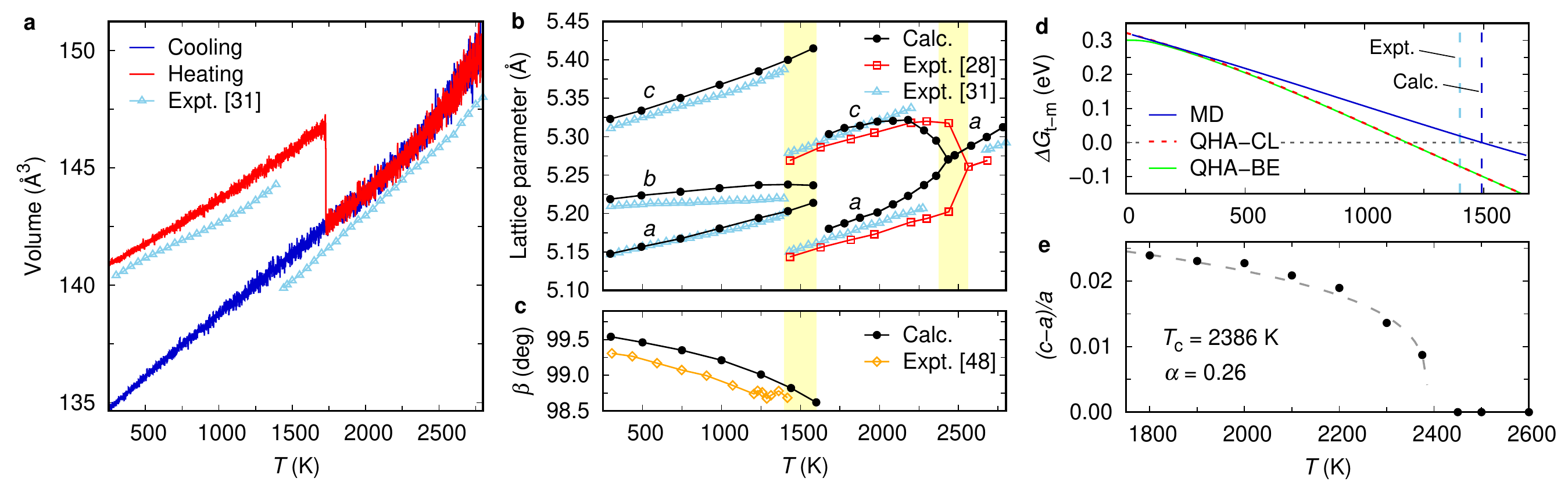}
 \caption{ \label{fig:phase-tr}
 \textbf{Phase transitions of ZrO$_2$ from MD simulations and thermodynamic
 integration.}
 \textbf{a} Evolution with temperature of the unit-cell volume during a
 heating and cooling simulation of a 324 atom supercell. \textbf{b-c}
 Temperature dependence of \textbf{b} the lattice parameters and \textbf{c}
 the monoclinic angle $\beta$ from NPT simulations. Experimental data from
 Refs.~\cite{Kisi1998,Aldebert1985,Subbarao1969} are also reported. In \textbf{a},
 the moving average over blocks of 2000 time steps is shown. The yellow areas in
 \textbf{b} and \textbf{c} highlight the difference between the experimental
 transition temperatures and the ones from direct MD simulations. The phase
 transitions are studied in more detail in \textbf{d}, showing the difference
 between the free energy per unit cell of the tetragonal and monoclinic phase
 $\Delta G_\textup{t-m}$ as a function of temperature, and \textbf{e}, displaying
 the tetragonal distortion $(c-a)/a$ and its fit to the function $(T_\textup{c}-
 T)^\alpha$. In \textbf{d} the data from our fully anharmonic MD calculations
 are reported, as well as from the QHA using classical (CL) and quantum (BE)
 statistics. The dashed vertical lines indicate the experimental and calculated
 temperature of the martensitic transformation. }
 \end{center} \end{figure*}

We now move to studying the transitions between the three different structural
phases of ZrO$_2$ at ambient pressure using MD simulations with our MLFFs. The
results shown are from the MLFF obtained using the SVD. The phase transitions can  be
directly observed by progressively heating a relatively large supercell containing
324 atoms. Specifically, we used a heating rate of 0.5~K ps$^{-1}$ and a time step of
1.5~fs in the isothermal-isobaric (NPT) ensemble. The evolution with temperature of the unit-cell volume
resulting from this simulation is illustrated in red in Fig.~\ref{fig:phase-tr}a.
As in the experimental data, reported as well, the first order transition between
the monoclinic and tetragonal phase corresponds to a sharp change in the cell
volume around 1750~K, while there is no volume discontinuity in the tetragonal
to cubic transformation around 2400~K, but only a small change in the slope of
the thermal expansion. An analogous cooling simulation showed that the
transformation between the cubic and tetragonal phase is reversible and presents
no thermal hysteresis, whereas the tetragonal to monoclinic transition was not
observed upon cooling. In Fig.~\ref{fig:phase-tr}b-c the lattice
parameters and monoclinic angle $\beta$ are reported, respectively. These
results were extracted from NPT simulations carried out at fixed temperature
for 300~ps, taking the equilibrium structure at each temperature as the starting
configuration for the next (higher) temperature point. Near the phase transitions,
the total time was increased to 750~ps due to the large fluctuations in the case
of the tetragonal to cubic transformation, and to the time required to observe
the first-order monoclinic to tetragonal transition. The comparison with available
experimental data shows excellent agreement, including the anisotropy of the
lattice thermal expansion in the monoclinic phase, however, further investigation
on the phase transitions is warranted.

To accurately determine the transition temperature we used thermodynamic
integration, as explained in this paragraph. As seen in
Fig.~\ref{fig:phase-tr}a-c, the martensitic transformation between the
monoclinic and tetragonal phases proceeds abruptly, and because of hysteresis
it is not possible to extract a reliable transition temperature from direct MD
simulations. For example, when heating the system at the rate of 0.5~K ps$^{-1}$ 
the transition appears around 1750~K, but when performing MD simulations at fixed
temperatures we could observe the transition already at 1700~K after 700~ps.
Moreover, this transition was not reversible when cooling the system. To estimate
the theoretical transition temperature, we therefore calculated the fully
anharmonic free energies of the monoclinic and tetragonal phase as a function
of temperature. From the thermodynamic relations in the NPT ensemble, the Gibbs
free energy $G$ of a system at constant pressure can be written as:
\begin{equation} \label{eq:free-ene}
 G(T)=-T\int_{T_0}^T \frac{H(T')}{T'^2}\,\mbox{d}T'+\frac{G_0}{T_0}T,
\end{equation}
where $H=U+PV$ is the enthalpy ($U$ is the internal energy of the system), and
$G_0$ is the Gibbs free energy at $T=T_0$. Since the integrand in the first term
on the r.h.s. diverges like $1/T'$ for classical statistics, direct integration
from $T_0=0$ is not possible. Instead, we here calculated $G_0$ at
a low temperature $T_0=25$~K using the quasi-harmonic approximation (QHA). Within
the QHA, the Helmholtz free energy $F$ of the crystal is computed as a function
of temperature $T$ and volume $V$ via the standard harmonic approximation, while
the Gibbs free energy is found by minimizing $F(V,T)+PV$ for a given temperature
and pressure~\cite{Togo2015}. Thus, the anharmonic effects are taken into account
only via the volume dependence of the vibrational frequencies. To be compatible
with the classical MD simulations, we computed the harmonic free energies using
classical Maxwell-Boltzmann statistics. We evaluated the integral in
Eq.~\eqref{eq:free-ene} over a path of constant (ambient) pressure for both
monoclinic and tetragonal ZrO$_2$ by performing additional MD simulations at
temperatures ranging from 25~K to 1700~K, sampling each trajectory for 300~ps.
The integration paths are continuous, since the tetragonal phase is trapped in
a local minimum and does not transform into the monoclinic one during our MD
simulations (compare Fig. \ref{fig:phase-tr}a), while the monoclinic phase
remains stable for sufficient time up to 200~K above the transition temperature.

The free energy difference $\Delta G_\textup{t-m}$ between the tetragonal
and monoclinic phase as a function of temperature is illustrated in
Fig.~\ref{fig:phase-tr}d. The results obtained from Eq.~\eqref{eq:free-ene}
are shown in blue and compared to the ones calculated using the QHA either in
the classical approximation (red line) or using quantum statistics (green line).
In the QHA, both statistics yield identical transition temperatures $T_\textup{c}$,
with the differences between both statistics being entirely negligible above 250~K.
This suggests that the neglect of quantum statistics is justified for free energy
calculations of ZrO$_2$. According to our fully anharmonic free energy
calculations, the transition temperature is 1493~K, in good agreement with the
experimental value of 1400~K. Above this temperature, the tetragonal phase is
thermodynamically stable due to the vibrational entropy. Instead, the QHA
underestimates the transition temperature by more than 200~K ($T_\textup{c}=
1178$~K), highlighting the need to describe anharmonicity beyond the QHA.
We note that the fully anharmonic calculations performed with the MLFF trained
using BLR yield $T_\textup{c}=1530$~K, in line with the slightly larger predicted
energy difference between the tetragonal and monoclinic phase at 0~K (see
%Table~\ref{tab:latt-par})
Table~2). The accuracy of our MLFFs in describing anharmonic
effects was further confirmed by comparing the QHA results obtained with the
MLFF and from first principles, yielding a difference of 20~K in the transition
temperature.

The tetragonal to cubic phase transition can be more easily described using
direct MD calculations; however, the nature of this transition is not clear
experimentally as cubic zirconia is observed only at temperatures roughly above
2570~K, making it difficult to study. From our MD data we observe a
continuous transformation with no thermal hysteresis, moreover the system
undergoes frequent fluctuations between the two phases near the transition
temperature. Our results therefore suggest that the transition is largely
second order, at variance with a similar calculation based on a semiempirical
interatomic potential~\cite{Schelling2001}. 
We note that due to the strongly anharmonic energy surfaces, the lattice 
parameters change rather quickly (but continuously) near the cubic to tetragonal  
transition temperature. As shown in Ref.~\cite{Finnis2001} on the basis of 
Landau theory, this phase transition is highly sensitive to the coupling between 
the elastic strains and the distortion of the oxygen sublattice. Hence, 
an interatomic potential must be able to describe the system quantitatively 
with high accuracy. This is not the case even for the `best' model interatomic 
potential for ZrO$_2$~\cite{Schelling2001}, whereas MLFFs can attain 
first-principles accuracy. Fitting the tetragonal distortion
$(c-a)/a$, as shown in Fig.~\ref{fig:phase-tr}e, yields a transition
temperature $T_\textup{c}$ of 2386~K, only 7\% lower than the experimental
one, and a critical exponent of 0.26 close to the value of 0.25 predicted by
Landau theory for a tricritical phase transition. Using BLR we obtain a very
similar value of 2384~K, not surprisingly as the soft phonon modes in the cubic
phase at 0~K are very accurately described by the BLR on par with SVD.
In contrast, previous theoretical studies based on a classical force
field~\cite{Schelling2001} and on a tight-binding model parametrized on
first-principles LDA calculations~\cite{Finnis2001} underestimated the
experimental transition temperature by almost 30\%.

 \begin{figure}[b] \begin{center}
 \includegraphics[width=0.5\columnwidth]{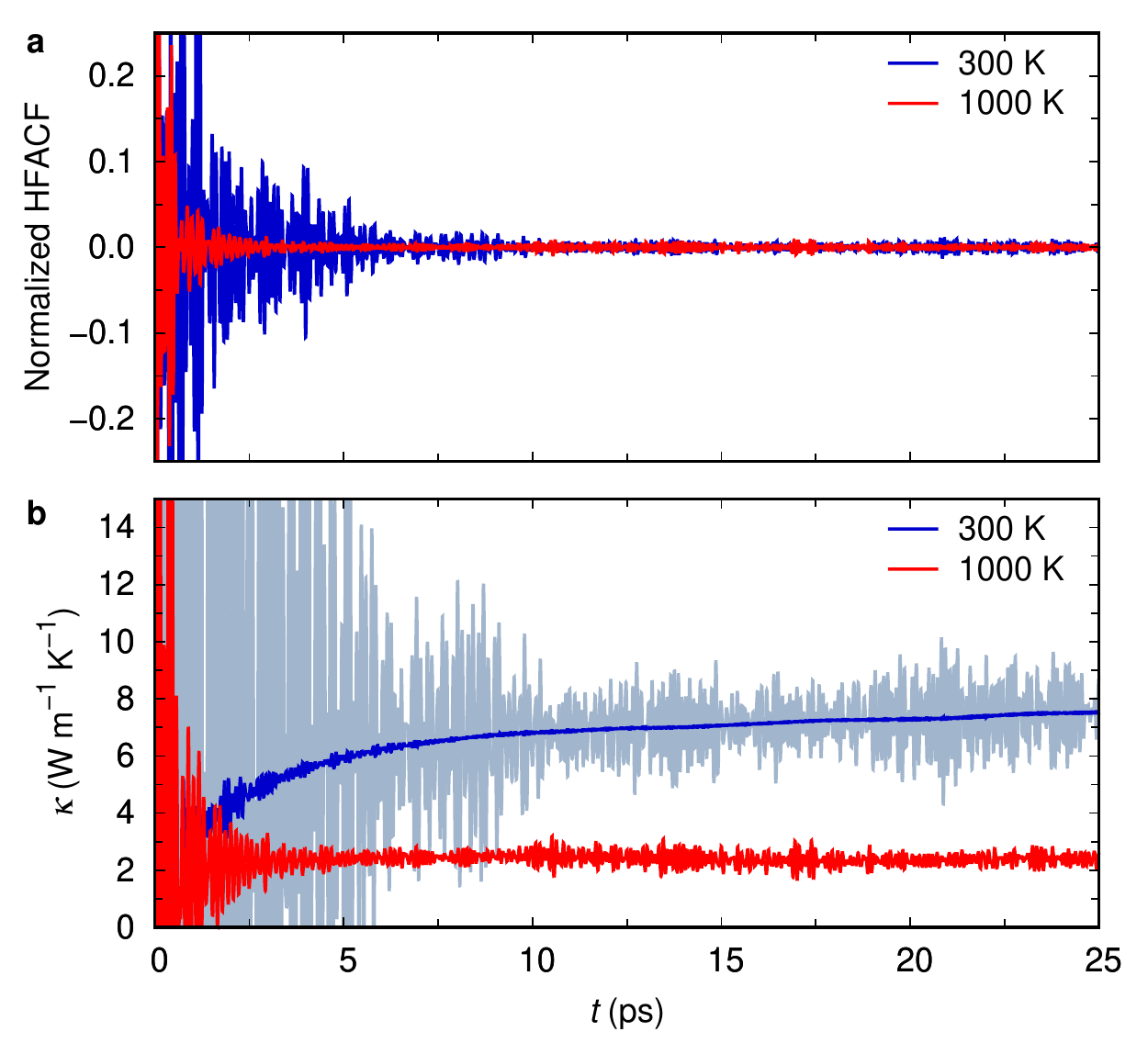}
 \caption{ \label{fig:heat-corr}
 \textbf{Calculating the thermal conductivity using GK theory.}
 \textbf{a} Heat-flux autocorrelation function (HFACF) normalized by its
 zero-time value and \textbf{b} thermal conductivity as a function of
 correlation time at $T=300$~K and 1000~K. In \textbf{b}, the moving average
 of $\kappa$ over blocks of 1000 time steps is shown on top of the raw data
 at $T=300$~K. }
 \end{center} \end{figure}

\subsection{Lattice thermal conductivity}

Finally, we used our MLFFs to compute the thermal conductivity $\kappa$ of
ZrO$_2$ using GK theory (see Methods). As pointed out in the
Introduction, this theory accounts for anharmonic scattering to all orders, and
it provides an exact description of heat transport as long as quantum statistical
effects are negligible. Zirconia is an interesting test case for the calculation
of the thermal conductivity combining MLFFs and GK theory as it is a strongly
anharmonic system. To calculate the heat flux [Eq.~\eqref{eq:heat2}], ZrO$_2$
supercells were first equilibrated in the NVT ensemble at the desired temperature
and equilibrium volume, taking into account the monoclinic to tetragonal
transformation around 1500~K, and the independent configurations were randomly
selected. For these configurations, the heat flux was then sampled during MD
simulations in the microcanonical ensemble using a time step of 1.5~fs. The
ensemble average in Eq.~\eqref{eq:kappa} was performed over 10 independent
trajectories, each up to 600~ps long. After testing the convergence of $\kappa$
with respect to the system size, supercells containing 324 atoms were selected
for all calculations.

As an illustration, the averaged heat-flux autocorrelation function (normalized
by its zero-time value) is shown in Fig.~\ref{fig:heat-corr}a as a function of
correlation time for two different values of the temperature. Note that it does
not decay monotonically but features large oscillations that are related to optical
phonons.
The corresponding cumulative time integrals that yield the thermal conductivities
are plotted in Fig.~\ref{fig:heat-corr}b. To better highlight the convergence of
$\kappa$ at 300~K with respect to the upper integration time, the moving average
is shown on top of the raw data. Using a standard procedure~\cite{Howell2012,
McGaughey2004}, the resulting value of $\kappa$ at each temperature was obtained
by averaging the integral over 1000 time steps in the region where it becomes
constant, before the statistical noise in the tail of the heat-flux autocorrelation
function dominates the integral (after 16~ps at 300~K).

 \begin{figure}[b] \begin{center}
 \includegraphics[width=0.5\columnwidth]{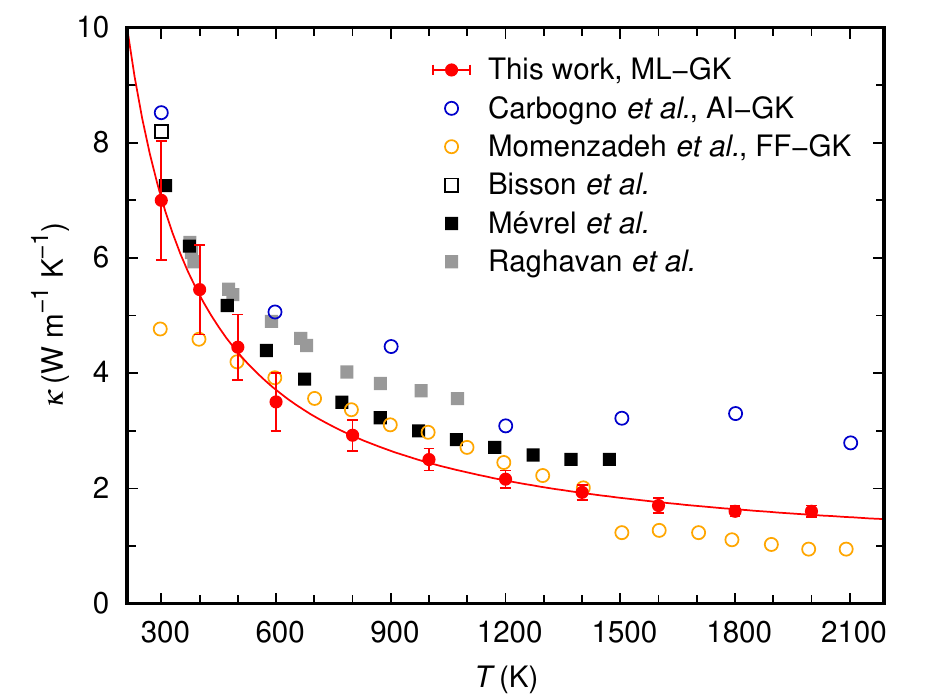}
 \caption{ \label{fig:kappa}
  \textbf{Thermal conductivity of ZrO$_2$ as a function of temperature.}
  The present results obtained using a MLFF and GK theory (ML-GK, red filled
  circles; the error bars are standard errors) are compared with previous theoretical results from GK theory
  combined with AI calculations (AI-GK, empty blue circles)~\cite{Carbogno2017}
  and an empirical force field (FF-GK, empty orange circles)~\cite{Momenzadeh2020}.
  Experimental data from Refs.~\cite{Bisson2000,Mevrel2004,Raghavan1998} are
  shown as square symbols. The red line is a fit to the present data. }
 \end{center} \end{figure}

The calculated values of $\kappa$ are shown in Fig.~\ref{fig:kappa} as isotropic
averages. In this case, BLR and SVD yield nearly the same results, therefore only
one set of data is shown. Our results are in reasonable agreement with the
experimental data for pure zirconia~\cite{Bisson2000,Mevrel2004}, available only
for the monoclinic phase, underestimating them by no more than 20~\%. As discussed
in Ref.~\cite{Mevrel2004}, it is unclear why the thermal conductivity measured in
Ref.~\cite{Raghavan1998} decreases less rapidly with temperature, but these data
are also reported in Fig.~\ref{fig:kappa} for completeness. Note that our
calculated $\kappa$ does not show any discernible discontinuity across the phase transition
temperature. Moreover, we found no significant anisotropy, in agreement with
experimental observations~\cite{Bisson2000}. Previous calculations based on GK
theory and using an empirical force field~\cite{Momenzadeh2020} are unable to
produce the correct temperature dependence and underestimate the room-temperature
thermal conductivity by 40\%. In contrast, the first-principles GK calculations
from Ref.~\cite{Carbogno2017} overestimate both our results and the experiment,
and show a less smooth temperature dependence which could indicate limited
convergence or limited statistical sampling. We note that in standard MD
simulations the dynamics of the nuclei is purely classical, which may yield higher
phonon scattering rates and thus lower thermal conductivities~\cite{Galli2012,
Galli2019}, although this effect should be negligible at sufficiently high
temperatures.

\section{Discussion}

In this work we machine-learned an interatomic potential for zirconia using 
an on-the-fly training scheme. We demonstrated that the MLFF allows to 
accurately describe the harmonic lattice dynamics and 
the anharmonic higher order contributions, with the anharmonic 
phonon-phonon interactions driving the temperature-dependent behavior, 
up to very high temperatures near the melting point. In particular, the solid-solid
phase transformations that are especially hard to study using AI methods were
correctly captured, and the transition temperatures were predicted with high
accuracy. Conversely, conventional force fields are not sufficiently 
flexible to achieve a quantitative description of the harmonic (quadratic) and 
higher-order energy terms. 
Likewise, modelling thermal conductivity is a stringent test of an accurate 
description of the anharmonic, high-order interatomic force constants terms. 
The same MLFF is indeed capable of predicting the lattice thermal
conductivity using GK theory, the gold standard for computing thermal
transport in strongly anharmonic crystalline and amorphous solids. This work 
demonstrates that machine-learned potentials have unlocked the ability to routinely 
perform GK calculations with near first-principles accuracy, as these calculations 
are generally prohibitive for AI methods and are beyond the accuracy of classical 
force fields.

It is noteworthy that although MLFFs cannot extrapolate well beyond the phase 
space of the training data, as is well known, we showed that a MLFF that is highly 
accurate up to almost 3000~K can be trained on a dataset consisting of as little as 
a few hundred structures using the present on-the-fly method. This 
method provides a general strategy for constructing an optimal training database 
via a semi-automated process. In the future, this can also be exploited for 
high-throughput predictions of vibrational and thermodynamic properties.
Interestingly, we demonstrated that the routinely used regularization is
not necessary in general in order to prevent overfitting when training a MLFF,
and an unregularized regression based on SVD systematically improves the accuracy
of the MLFF. As showcased in this work, on-the-fly machine-learned interatomic
potentials are an enabling tool for accurate and efficient simulations of complex
thermodynamic properties of a wide variety of technologically relevant materials.
%\vspace{20pt}

%%% METHODS <3000 words (with subheadings)
\section{Methods}

\subsection{On-the-fly machine-learned force fields} \label{sec:MLFF}

To build a machine-learned interatomic potential, first of all, a set of
reference configurations with their quantum-mechanical energies, atomic
forces and, ideally, stress tensors is required. Active learning schemes
provide an extremely efficient solution for constructing such a database,
whereby automated query strategies allow the machine to judge whether new
structures should be added to the training dataset or not~\cite{Jinnouchi2020r}.
Extensive details of the algorithm as well as of the machine-learning model
can be found in Refs.~\cite{Jinnouchi2019b,Jinnouchi2020}, while here only
the most relevant features are outlined. The core of our training strategy
is that the MLFF is constructed on the fly during MD simulations, 
and at every MD step it is used
to judge whether a first-principles calculation should be executed and a new
structure should be added to the dataset. The decision relies on an estimation
of the errors in the predicted atomic forces for each structure on the basis
of Bayesian inference, as it will be explained later. When no new 
first-principles calculations are carried out, the atomic positions and velocities 
are updated using the MLFF predictions.

Machine-learning models for building interatomic potentials rely on a local
mapping of structural features onto a set of descriptors. A robust and versatile
choice that allows for an accurate description of the many-body interactions is
to map the local environment around each atom $i$ in a system of $N_\textup{a}$ 
atoms onto two- and three-body atomic distribution functions~\cite{Bartok2013}. 
In this work we relied on so-called separable descriptors introduced in 
Ref.~\cite{Jinnouchi2020}, with the two- and three-body distribution 
functions ($\rho_i^{(2)}$ and $\rho_i^{(3)}$, respectively) defined as:
 \begin{gather}
 \rho_i^{(2)}(r) = \:\frac{1}{4\pi} \int \!\!d\hat\br \,\rho_i(r\hat\br)\,; \\[4pt]
 \rho_i^{(3)}(r,s,\theta) = \int\!\!\!\!\int \!\!d\hat\br d\hat{\mathbf{s}} 
 \:\delta(\hat\br\cdot\hat{\mathbf{s}} -\cos\theta) 
 \times \!\bigg[ \rho_i(r\hat\br) \rho_i^*(s\hat{\mathbf{s}}) 
  -\sum_{j\neq i}^{N_{\textup{a}}} \tilde\rho_{ij}(r\hat\br) 
  \tilde\rho^*_{ij}(s\hat{\mathbf{s}}) \bigg].
 \end{gather}
In these expressions, $\rho_i(\br)$ ($\br=r\hat\br)$ is the three-dimensional  distribution function around the atom $i$, $\rho_i(\br)=\sum_{j\neq i}^{N_{\textup{a}}} 
\tilde\rho_{ij}(\br)$, and $\tilde\rho_{ij}(\br)$ is the likelihood to find 
atom $j$ at position $\br$ from atom $i$ within a certain cutoff radius. 
Essentially, $\rho_i^{(2)}(r)$ yields the probability to find an atom $j$ ($j\ne i$) 
at a distance $r$ from atom $i$, and $\rho_i^{(3)}(r,s,\theta)$ the probability 
to find two atoms $j,k$ at a distance $r,s$, respectively, from atom $i$ ($j\ne i$, 
$k\ne j,i$) and spanning the angle $\theta$ between them. 
In practice, these distribution functions are expressed as a combination of 
radial and angular basis functions, with the coefficients forming a set of 
descriptors that we simply denote as $\mathbf x_i^{(2)}$ and 
$\mathbf x_i^{(3)}$. Following \cite{Jinnouchi2020}, we define $\mathbf x_i=
(\mathbf x_i^{(2)},\mathbf x_i^{(3)})$.

The quantum-mechanical energies, atomic forces and stress tensors can then be
expressed as a nonlinear function of the descriptors. In the present method, the
potential energy $U$ of a system of $N_\textup{a}$ atoms is partitioned into a
sum of local energies $U_i$ associated to each atom. Following the Gaussian
approximation potential (GAP) approach~\cite{Bartok2010}, each term $U_i$ is
written as a linear combination of kernel functions $K\left(\mathbf x_i,
\mathbf x_{b} \right)$ weighted by coefficients $w_{b}$:
\begin{equation} \label{eq:en-GAP}
 U = \sum_{i=1}^{N_\textup{a}} U_i = \sum_{i=1}^{N_\textup{a}} \sum_{b=1}^{N_b}
 w_{b}\, K\left(\mathbf x_i,\mathbf x_{b} \right).
\end{equation}
In practice, the kernel function measures the similarity between a local
configuration around atom $i$, $\mathbf x_i$, and a local reference configuration
around atom $b$, $\mathbf x_{b}$. The $N_b$ atoms are chosen from the set
of reference structures generated for the MLFF training. Here $K$ is the 
dot-product kernel, determined by $K\left(\mathbf x_i,\mathbf x_{b} \right)=
\left( \mathbf{\hat x}_i \cdot \mathbf{\hat x}_{b} \right)^\zeta$, with
$\mathbf{\hat x}$ the normalized descriptors. The normalization and exponentiation
by $\zeta$ introduce non-linear terms that mix the two- and three-body descriptors
and generate many-body terms necessary for capturing higher order atomic
interactions.

By the same token, the forces on each atom and the stress tensor components
can also be written in terms of the kernel functions. Then, the coefficients
$\mathbf w=\{w_{b}\}$ are optimized to best reproduce the first-principles
energies per atom, forces and stress components in the training dataset, in
other words by solving the problem:
\begin{equation} \label{eq:w-lsq}
 \mathbf y = \bm\Phi\,\mathbf w \quad \rightarrow \quad
 ||\mathbf y - \bm\Phi\,\mathbf w || = \mbox{min}
\end{equation}
in a least-squares sense, where $\mathbf y$ is a vector collecting all the
first-principles data, and $\bm\Phi$ is the so-called design matrix containing
the $K(\mathbf x_i,\mathbf x_{b})$ components and the derivatives with respect
to atomic coordinates and lattice vectors. When fitting simultaneously energies,
forces and stress tensors, as done here, it is expedient to rescale these
quantities by their respective standard deviations~\cite{Jinnouchi2019b}. To
increase the accuracy of the MLFF, it is also possible to tune the relative
importance of energies, forces and stresses during the fitting.

The solution of the linear regression problem in Eq.~\eqref{eq:w-lsq} is a
crucial step. The standard methods used for training MLFFs are based on ridge
regression, where a regularization parameter $\sigma_\textup{v}^2/
\sigma_\textup{w}^2$ is introduced~\cite{Bishop}. In the probabilistic 
framework of Bayesian linear regression (BLR) adopted here, the coefficients 
are formally determined in the same way:
\begin{equation} \label{eq:w-BLR}
 \mathbf w = \left( \mathbf\Phi^\textup{T}\mathbf\Phi+ \sigma_\textup{v}^2/
 \sigma_\textup{w}^2\, \mathbf I \right)^{-1} \mathbf\Phi^\textup{T} \mathbf y,
\end{equation}
however, here the regularization parameters $\sigma_\textup{v}$ and 
$\sigma_\textup{w}$ have a probabilistic meaning: $\sigma_\textup{v}^2$ models 
the uncertainty caused by noise in the training data, while $\sigma_\textup{w}^2$ 
is the variance of the prior distribution of the regression coefficients. These values 
are optimized using the evidence approximation~\cite{Bishop,Jinnouchi2019b} and 
are generally invoked to prevent overfitting. Moreover, the 
probabilistic framework of BLR is advantageous for the on-the-fly training. 
In fact, under the assumption that the prior distribution of the target data (that is, 
the \textit{ab~initio} energies, forces and stress tensors) follows a multivariate 
Gaussian distribution, from Bayesian inference it derives that the posterior 
distribution of the predicted data is a Gaussian distribution too. The BLR 
framework then allows us to define a `prediction variance' that measures the 
uncertainty of the predictions. For any given configuration, 
this is estimated as the variance of the posterior distribution: 
\begin{equation}
\bm\sigma^2 = \sigma_\textup{v}^2 \,\mathbf I +\sigma_\textup{v}^2 \,\bm\varphi 
\,\Big( \bm\Phi^T \bm\Phi+\sigma_\textup{v}^2/ \sigma_\textup{w}^2 \,
\mathbf I \Big)^{-1} \bm\varphi^T,
\end{equation}
where $\bm\varphi$ is a matrix containing the kernel components and their partial 
derivatives with respect to atomic coordinates and lattice vectors for the given 
configuration. This is the uncertainty estimate that guides the selection of the 
training dataset during on-the-fly active learning.

Alternatively, in this work the regression coefficients were also 
determined as:
\begin{equation} \label{eq:w-SVD}
 \mathbf w =\mathbf\Phi^+\mathbf y ,
\end{equation}
where $\mathbf\Phi^+$ denotes the Moore-Penrose pseudoinverse of the design matrix
$\mathbf\Phi$, which can be found by performing the singular-value decomposition
(SVD) of $\mathbf\Phi$. Even though SVD carries a larger computational cost, a
clear advantage of this method is that the condition number of the problem is
smaller, since the design matrix is not squared. As in Ref.~\cite{Liu2021}, we
found that while the squared problem in Eq.~\eqref{eq:w-BLR} is generally
ill-conditioned and requires regularization, Eq.~\eqref{eq:w-SVD} can be solved
numerically without regularizing the singular values of the matrix $\mathbf\Phi$.
As our tests show, this does not lead to overfitting, demonstrating that
regularization is not necessary in general. As this is a deterministic 
method, SVD is performed only at the end of the on-the-fly training process, and 
the different MLFFs obtained by using BLR and SVD can then be compared.

\subsection{Thermal conductivity via Green-Kubo theory} \label{sec:GK}

On the basis of GK theory, the thermal conductivity $\kappa$ is related to the
equilibrium average of the heat-flux autocorrelation function~\cite{Baroni2018}:
\begin{equation} \label{eq:kappa}
 \kappa_{\alpha\beta}=\lim_{t\to\infty}\frac{1}{\kb T^2V}
 \int_0^t \braket{j_\alpha(t') j_\beta(0)} \mbox{d}t',
\end{equation}
where $\kb$ is the Boltzmann constant, $T$ the temperature, $V$ the volume of the
system and $j_\alpha(t)$ the $\alpha$-th Cartesian component of the macroscopic
heat flux. The symbol $\braket{\cdot}$ denotes an ensemble average over many time
origins and over independent MD trajectories.

The heat flux for a system of $N_\textup{a}$ atoms is given by~\cite{Hardy1963}:
\begin{equation} \label{eq:heat1}
 \bj(t)=\sum_{i=1}^{N_\textup{a}}\frac{d}{dt}(\br_i E_i),
\end{equation}
where $E_i=m_i \bv_i^2/2+U_i$ is the total (kinetic and potential) energy
associated to atom $i$ with mass $m_i$, while $\br_i$ and $\bv_i$ are its position
and velocity, respectively. For classical many-body potentials, this definition
does not pose any problem, since the total energy of the system is naturally
partitioned into on-site contributions $E_i$. This holds for machine-learned
potentials too, as the potential energy $U$ is expressed in terms of the
potential energies associated to each atom $U_i$ [cfr. Eq.~\eqref{eq:en-GAP}].
Eq.~\eqref{eq:heat1} is readily rewritten in a form applicable to periodic
systems~\cite{Harju2015}:
\begin{equation} \label{eq:heat2}
 \bj(t)=\sum_{i=1}^{N_\textup{a}}\bv_i E_i + \sum_{i=1}^{N_\textup{a}}
 \sum_{j\neq i}^{N(i)} (\br_i-\br_j)\left(\frac{\partial U_i}{\partial \br_j}
 \cdot\bv_j\right),
\end{equation}
where the summation over $j$ is carried out over all $N(i)$ atoms within the
cutoff radius representing the local environment of atom $i$. The first term
in Eq.~\eqref{eq:heat2} is the convective heat flux, which gives no contribution
to the thermal conductivity in solids, while the second one is the so-called
virial or conductive heat flux.

Combining Eqs.~\eqref{eq:kappa} and \eqref{eq:heat2} it is thus possible to
evaluate the thermal conductivity from the atomic energies provided by MLFFs
in a similar manner as when using classical force fields. The advantage here
is that the AI description of the energy landscape is retained.

\subsection{First-principles calculations and MLFF training} \label{sec:comp-det}
%\subsection{Computational details and MLFF training} \label{sec:comp-det}

All calculations were performed using VASP. First-principles DFT calculations
were carried out using the PBEsol~\cite{Perdew2008} exchange-correlation
functional. The plane-wave cutoff was set to 600~eV. Phonon calculations were
performed using the Phonopy package~\cite{Togo2015}. The primitive cells of
zirconia contain one, two, and four ZrO$_2$ formula units for the cubic,
tetragonal and monoclinic structures, respectively. A unit cell of 12 atoms
can therefore be used to describe all three phases, and a $4\times4\times4$
Monkhorst-Pack $k$-mesh ensures energy convergence within less than 0.2~meV 
per atom.
The structures were optimized until the forces were smaller than $10^{-4}$~eV 
\AA$^{-1}$.

Our MLFF for solid ZrO$_2$ was trained on the fly during MD simulations in the
NPT ensemble at ambient pressure, with the temperature and pressure controlled
by a Langevin thermostat~\cite{Allen-Tildesley} combined with the Parrinello-Raman
method~\cite{Parrinello1980,Parrinello1981}. The training was performed on a
supercell containing 96 atoms using a time step of 2.5~fs. First, the system
was gradually heated from 1600~K to 2800~K for 180~ps, then it was heated again
starting from 500~K to 1600~K for another 180~ps. The initial structures for
these MD simulations were obtained by equilibrating the tetragonal and monoclinic
structures at 1600~K and 500~K, respectively, for 50~ps using the on-the-fly
scheme and discarding the MLFFs thus generated. Finally, to sample the
low-temperature, unstable cubic, and tetragonal structures as well, we performed 10
additional MD steps at 100~K starting from the DFT-relaxed structures. In total,
only 592 first-principles calculations were executed, out of almost 145,000 MD
steps. These constitute the reference structures dataset, from which $N_b=831$
and 1090 local reference configurations were selected for Zr and O, respectively.

The main hyperparameters used for constructing the descriptors and kernel
functions~\cite{Jinnouchi2019b,Jinnouchi2020} are as follows. The cutoff radius
used to represent the local environment of each atom was set to 6~\AA, while
the width of the Gaussian functions used to broaden the atomic distributions was
0.4~\AA. These parameters were the same for the two- and three-body descriptors.
To expand the atomic distributions, up to 15 spherical Bessel functions (for
the angular quantum number $l=0$) and Legendre polynomials of order $l$ up to
$l_\textup{max}=4$ were employed, resulting in 2048 descriptors for 
the Zr and O atoms. The hyperparameter for the dot-product kernel
was set to the standard value $\zeta=4$. The automatic sparsification of the
three-body descriptors based on the CUR algorithm was applied~\cite{Jinnouchi2020},
allowing us to halve the number of descriptors needed without affecting the
accuracy of the MLFF. As described previously, the MLFF can be further
tuned by reweighing the energies, forces and stresses differently when setting
up the linear regression problem. Here we found that weighing the energy equations
10 times more strongly reduces the errors in the energy by about 0.5~meV per atom
on average, while increasing the errors in the predicted forces and stresses
only marginally. Finally, both BLR and SVD were adopted to determine the
regression coefficients.

\section*{Data Availability}
The data that support the findings of this study are available from the
corresponding author upon reasonable request.

\section*{Code Availability}
The VASP code is distributed by the VASP Software GmbH. The machine learning 
modules will be included in the release of vasp.6.3. Prerelease versions are 
available from G. K. upon reasonable request.

\begin{acknowledgments}
The computational results presented here have been mainly achieved using the Vienna Scientific Cluster (VSC). This work was supported by the Austrian Science Fund FWF 
(SFB TACO). P.L. gratefully acknowledges the support of the Advanced Materials Simulation Engineering Tool (AMSET) project, sponsored by the US Naval Nuclear Laboratory (NNL) and directed by Materials Design, Inc. 
\end{acknowledgments}

\end{document}